# Insights into image contrast from dislocations in ADF-STEM

E. Oveisi[1,2]*, M.C. Spadaro[1], E. Rotunno[3], V. Grillo[3,4], C. Hébert[1,2]*

[1] *Interdisciplinary Centre for Electron Microscopy, École Polytechnique Fédérale de Lausanne (CIME-EPFL), Lausanne, Switzerland*

[2] *Electron Spectrometry and Microscopy Laboratory, École Polytechnique Fédérale de Lausanne (LSME-EPFL), Lausanne, Switzerland*

[3] *Institute of Nanoscience, National Research Council (NANO-CNR), Modena, Italy*

[4] *Institute of Materials for Electronics and Magnetism, National Research Council (IMEM-CNR), Parma, Italy*

**Abstract**

Competitive mechanisms contribute to image contrast from dislocations in annular dark-field scanning transmission electron microscopy (ADF-STEM). A clear theoretical understanding of the mechanisms underlying the ADF-STEM contrast is therefore essential for correct interpretation of dislocation images. This paper reports on a systematic study of the ADF-STEM contrast from dislocations in a GaN specimen, both experimentally and computationally. Systematic experimental ADF-STEM images of the edge-character dislocations revealed a number of characteristic contrast features that are shown to depend on both the angular detection range and specific position of the dislocation in the sample. A theoretical model based on electron channelling and Bloch-wave scattering theories, supported by multislice simulations using Grillo's strain-channelling equation, is proposed to elucidate the physical origin of such complex contrast phenomena.

---

Corresponding authors.
Email addresses: emad.oveisi@epfl.ch (E. Oveisi), cecile.hebert@epfl.ch (C. Hébert)

# 1. Introduction

Defect analysis in crystalline materials provides important insights into many properties of materials across a broad range of applications [1]. While conventional transmission electron microscopy (CTEM) has commonly been used to advancing the field of crystalline defect analysis, recently interest has emerged in using scanning TEM (STEM) to elucidate defect structure. STEM is particularly appropriate for studying much thicker samples than CTEM normally allows, as well as for obtaining images suitable for stereoscopic or tomographic reconstructions in which dynamical contrast effects are reduced [2-9]. Such interesting characteristics of the STEM images arise because the scattered state wave vectors are integrated over the acceptance range of an annular detector [10]. By using different detectors at the same time, various imaging modes, for instance bright-field (BF) and annular dark-field (ADF), are simultaneously accessible [11]. Depending on the detector geometry and beam convergence angle, there are several independent mechanisms contributing to defect contrast in the STEM images, making image interpretation complex [12, 13].

Since the first demonstrations of using STEM for defect analysis in the 1970s, many mechanisms have been proposed to treat the ADF-STEM images [10, 12, 14-18]. Maher and Joy were the first to apply the principle of reciprocity for defect image interpretation by using fixed-beam dynamical theory of electron diffraction [15]. They showed that the geometry of crystalline defects could be assessed by STEM diffraction analysis methods as in CTEM. Electron diffraction contrast has been used in many other studies to interpret defect image contrast [4, 19, 20]. Perovic *et al.* used Bloch-wave scattering theory as an alternative approach to elucidate the effect of elastic strain on the contrast of the ADF images [10, 18]. According to their theory, the contrast is associated with the Bloch-wave interference effects through the foil thickness, and therefore depends on the specific position of the defect in the foil. In the Bloch-wave theory of defect analysis, the presence of strain field affects the excitation of the Bloch-waves, resulting in inter-band transition between the Bloch-waves states. The Bloch-wave theory has also been applied to analyse the contrast around defects at the atomic scale [21].

A closely related approach for interpreting dislocation images contrast, known as de-channelling, was suggested by Cowley and Huang [14]. In the electron channelling process, the atomic sites act as nano-lenses concentrating the beam along the attractive potential of the column, which in the Bloch-wave model is considered as strong excitation of the localized *1s*-type waves [22-25]. The non-dispersive *1s* Bloch states that are localized on the atomic sites dominate the signal at high scattering angles [22, 23, 26]. In the de-channelling theory, image contrast results from any disruption of the channelling process. Distortion of the lattice channels by the strain field close to a crystal defect or, on a broader scale, loss of the wave-field symmetry close to crystal surface interrupt the channelling condition, thus can be considered as an origin for signal in the ADF images [14].

Grillo *et al.* studied the effect of long-range strain and local static displacements on the forward propagation of the wave function [17]. They proposed that in case of large static displacements the curvature of atomic planes should be taken into account instead of atomic displacements [17]. Where atomic column inclination happens, the wave states at either sides of the crystal needs to be matched, causing in turn an additional diffuse scattering. Local lattice distortion or atomic displacements at dislocation core also generates static diffuse scattering that, by analogy to X-Ray scattering process, is called Huang scattering [27-30]. Huang scattering analogous with the TDS has the effect of weakening the Bragg reflections but in contrary to the TDS is time independent and contains little information about the atomic number [30]. Grillo *et al.* also described the scattering angle dependence of the ADF strain contrast as a result of contribution of two competitive effects: Huang scattering and de-channelling [17]. Simulations demonstrated that strain fields reduces the propagation of the forward wave function, thus has a similar contribution to a wide range of scattering angles, from low to high angles [31]. On the other hand, the intensity of the scattering due to static displacements is angle-dependent, and decreases rapidly as the angle is increased. Therefore, at low angles, the contrast of static displacements is determined by the Huang scattering and de-channelling events, which act inversely. At higher angles, the contribution of Huang scattering becomes less important, and therefore contrast from strained region decreases as it is dominated by the de-channelling process [31].

Overall, due to the complexity of the mechanisms contributing to contrast from defects in ADF-STEM images, it is often essential to couple image simulations with experimental data for a correct interpretation of defect contrast [16, 19, 20]. Here, we use Bloch-wave scattering and electron-channelling theories to further report on the ADF-STEM dislocation contrast mechanism. We apply this approach to interpret the contrast effects that ascertained to depend on both ADF detection range and specific position of the defect in the sample. We couple the experimental images with simulations that have been done using an improved version of Grillo's strain-channelling equation system to further explore the depth-dependent ADF dislocation contrast [32, 33].

## 2. Experimental and computational procedures

### 2.1. Specimen

For this study, a thin film of gallium nitride (GaN) with relatively high content (about $10^{10}$ cm$^2$) of edge- and mixed-character dislocations was used (details on GaN growth in [34]). To prepare sample for STEM analysis, a cross-sectional lamella was extracted along the growth direction of the GaN membrane (i.e. parallel to its $(1\bar{1}0\,0)$ crystallographic plane). This process was done trough conventional focused ion beam (FIB) lift-out in a Zeiss NVision40. The thinning process was followed by low energy final polishing in the FIB (2 kV and 60 pA) to minimize the ion-induced damage and to obtain a foil with a uniform thickness of about 200 nm. During the preparation process the site of interest had been covered with a protective layer through FIB-assisted carbon deposition.

### 2.2. Electron microscopy

To investigate the dependence of image contrast from dislocations of deferent depths on the ADF-STEM detection, a systematic experiment varying the annular detector collection range was performed in a FEI Tecnai-Osiris at 200 kV. The contrast was studied for threading edge-character dislocations that were located close to entrant and exit foil surfaces while the specimen was tilted to satisfy a $[1\,\bar{1}\,0\,0]$ zone axis orientation. Burgers vectors of the threading dislocations in the GaN membrane were determined using the invisibility criterion (i.e. $\mathbf{g}\cdot\mathbf{b}$ analysis) under $\mathbf{g} = (0\,0\,0\,2)$ and $\mathbf{g} = (1\,1\,\bar{2}\,0)$ diffraction conditions (Supplementary Figure S1). The approximate depth of dislocations in the GaN foil was determined using thickness fringes (see Supplementary Figure S2) [35].
Series of STEM images were acquired with an ADF detector over a range of camera lengths (between 34 mm and 770 mm). Table 1 lists the collection range of the detector as a function of camera length. Images were acquired at two magnifications: low magnification, (1024x1024 pixel resolution, 2.42 nm pixel size, 48 msec dwell time); high magnification, (2048x2048 pixel resolution, 2.97 Å pixel size, 6 msec dwell time). At each camera length the detector gain (brightness and contrast) has been adjusted to increase the visibility of the image features. During the experiments, the illumination convergence semi-angle of the electrons was set to 9.7 mrad, using a nominal 50 µm condenser aperture. Energy filtered convergent beam electron diffraction (CBED) patterns were acquired on a JEOL-2200FS operated at 200 kV in nano-beam mode and zero-loss filtered with an energy slit of 30 eV.

Table 1. ADF-STEM detector collection angles (mrad) as a function of camera length (mm)[1].

|   | Collection angle (mrad) | |
|---|---|---|
|   | $\beta_{in}$ | $\beta_{out}$ |
| Camera length |   |   |
| 34 | 180 | 200 |
| 43 | 175 | 200 |
| 54 | 139 | 200 |
| 75 | 100 | 200 |
| 87 | 86.4 | 200 |
| 115 | 65.3 | 200 |
| 165 | 45.5 | 200 |
| 220 | 34.2 | 200 |
| 330 | 22.8 | 139 |
| 550 | 13.7 | 83.6 |
| 770 | 9.7 | 59.7 |

## 2.3. Numerical simulations

Grillo's strain-channelling equation has been implemented for the simulations of the ADF dislocation contrast. Grillo's strain-channelling equation starts from the Bloch-wave scattering theory and explicitly describes the evolution of the highly exited *1s* states after interaction with the sample [33, 36]. In order to introduce and take into account all sample features the equation contains several parameters: some of those depend on the sample and strain features (thickness, defect depth, strain related atomic column curvature, etc.), and others govern the contrast in different detection regimes. Here, we used a modified version of this code that more specifically describes the contribution of Huang scattering to image formation, in particular the scattering produced at middle angle range by higher energy non-1s Bloch states. These modifications are described in detail in the appendix. For simulations the defect core size is set to 25 Å, for a straight defect at different depths (from 20 to 180 nm) in a 200 nm thick GaN sample. During the simulations we included the surface relaxation. The material absorption coefficient and the curvature of the atomic column were fixed to $5\times10^{-3}$ Å$^{-1}$ and $3\times10^{-4}$ Å$^{-1}$, respectively. The source code is written in C$^{++}$ and has been implemented in the STEM_CELL software [33, 36].

---

[1] The upper limit of 200 mrad is due to the cut-off by the pole-pieces and inner tube of the microscope.

## 3. Results and discussions

### 3.1. Experimental results

The angular collection range β of an ADF detector ranges from low-angle at large camera lengths to medium-angle, and finally to high-angle when further decreasing the camera length. Based on the detectors' angular collection range, we classify the ADF images into three regimes: low-angle ADF (LAADF), 20 < β < 60 mrad; medium-angle ADF (MAADF), 40 < β < 120 mrad; and high-angle ADF (HAADF), $β_{in}$ > 80 mrad. In the LAADF regime, both elastic Bragg and diffuse scattering significantly contribute to image signal. Contribution of the Bragg scattering decreases in the MAADF regime, and in the HAADF regime it is overtaken by thermal diffuse scattering (TDS) and Rutherford scattering.

Overview ADF-STEM images of the sample at different regimes are shown in Figure 1. Different image contrasts are observed for the top- and bottom-configuration dislocations (i.e. close to entrant and exit foil surface, respectively) in different regimes, revealing that the ADF dislocation contrast depends sensitively on the specific position of a dislocation in the foil. The bottom-configuration dislocations gradually lose their visibility as the detector collection range is increased to MAADF regime, and finally are faintly visible in the HAADF regime. In contrary, the top-configuration dislocations remain visible in all the three regimes. Further investigations indicated that the contrast features also change between different regimes. To better visualize the evolution of dislocation contrast as a function of angular collection range, series of images from two dislocations that are located close to the entrant and exit foil surfaces (marked 1 and 2 in Figure 1, respectively) are presented together in Figure 2. The intensity profiles across these two dislocations are plotted in Figure 3. It highlights a contrast reversal for the top-configuration dislocations between the LAADF and HAADF regimes (see the region marked by red rectangles in Figure 2). The bottom-configuration dislocations however do not show this contrast reversal, but gradually lose their visibility towards the HAADF regime as the detector angular range is increased. To verify the reproducibility of such depth-dependent contrast mechanism, the aforementioned dislocations were imaged after swapping the specimen in the microscope. This is demonstrated in Figure 2b,c, confirming the existence of a similar trend. It should be noted that a similar trend has also been observed in the images that were simultaneously acquired with another ADF detector ("DF4") that has a smaller inner/outer diameter [37].

To further study the contrast reversal mechanism that happens for the top-configuration dislocations, images at higher magnification were acquired. The images highlight that the contrast features from a top-configuration dislocation totally change between different regimes. In the LAADF regime the contrast is composed of a narrow line (FWHM ≈ 4 nm) of negative contrast (i.e. with lower intensities relative to the background), surrounded by two parallel lines of positive contrast. The width of the region with positive contrast either sides of the dislocation line is about 7 nm. As of now in the text this contrast feature will be referred to as *M-type* contrast. As demonstrated in Figures 2 and 3b, such contrast however appears to be independent from dislocation position in the foil and is identical for both top- and bottom-configuration dislocations. The contrast of the top-configuration dislocations changes from the *M-type* at the LAADF regime to a sharp single peak, so called *I-type*, at the MAADF regime. This sharp peak remains visible for larger collection ranges. Finally, at the HAADF regime, the top-configuration dislocations exhibit a so-called *W-type* contrast; a contrast that is complementary to the *M-type* contrast of the LAADF regime. An example of the *M-*, *I-,* and *W-type* contrasts are illustrated in Figure 4.

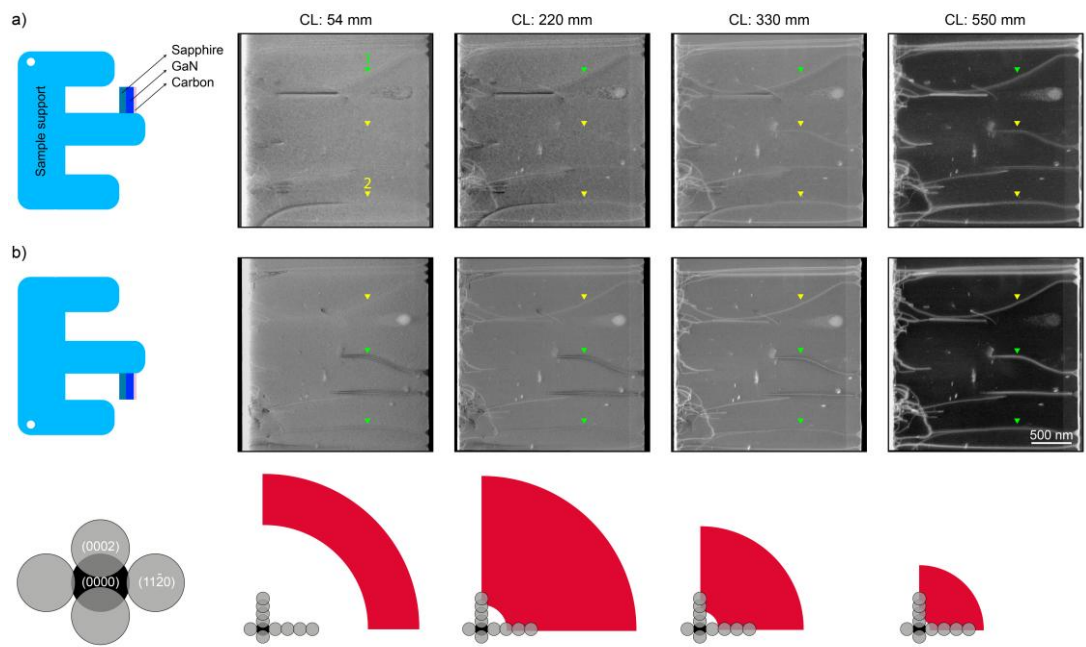

Figure 1. A comparison between dislocation contrasts over a range of ADF detection angles. (a) Changes in the contrast from dislocations of different specific depths in the GaN sample with varying the ADF detection range. Arrows encode the approximate position of dislocation in the specimen: green, dislocation close to entrant foil surface; yellow, dislocation close to exit foil surface. The GaN sample was orientated to have the $[1\bar{1}00]$ direction parallel with the electron beam. (b) ADF-STEM images of the same region of the sample after flipping the specimen upside down in the microscope. To ease the comparison, the bottom row images (b) are flipped horizontally. To the bottom of each column, a schematic of the ADF detector and zone axis STEM diffraction configuration are shown. Schematics on the left side of each row demonstrate the geometry of the specimen in the microscope.

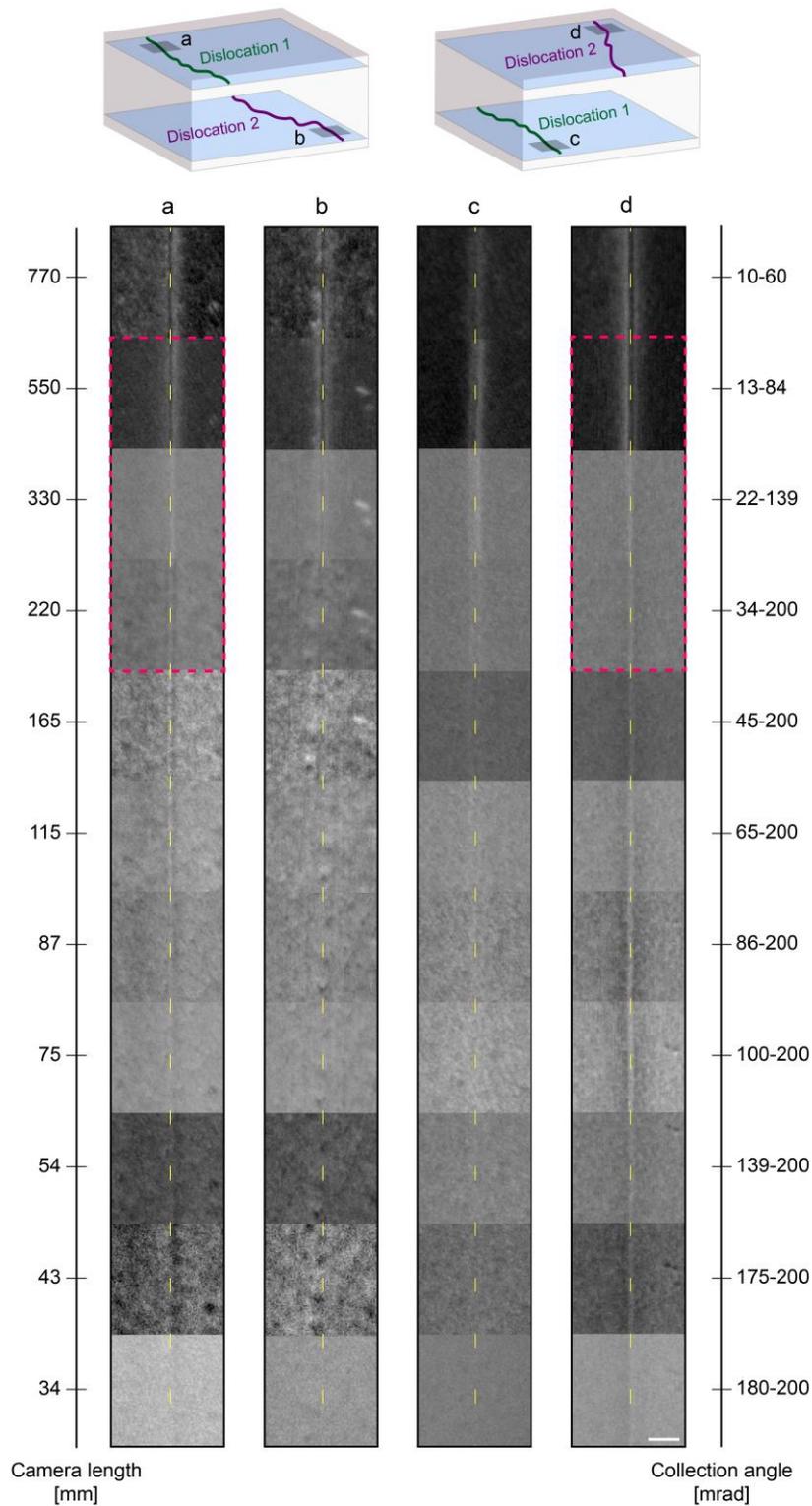

Figure 2 – Illustration of dislocation image contrast evolution with varying the ADF detection range. (a,b) Dislocations 1 and 2 of Figure 1, respectively. Images of the same dislocations after flipping the specimen upside down in the microscope are shown in (c, close to exit foil surface) and (d, close to entrant foil surface). The change in contrast is different for the top- and bottom-configuration dislocations. The contrast has been adjusted to increase the visibility of the image features. Scale bar is 50 nm.

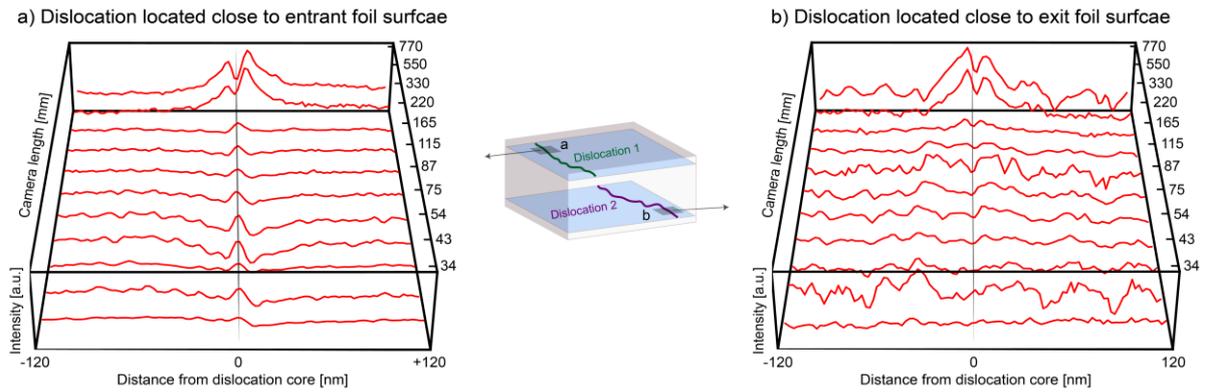

Figure 3 – Line scans of intensity profile across dislocations of Figures 2a and 2b. (a) Dislocation 1 (of Figure 1), located close to entrant foil surface; b) Dislocation 2, located close to exit foil surface. The transparent grey plane on the plots indicates the approximate position of the dislocation core. The intensity profiles are 240 nm long and the intensity is integrated over a region of 60 nm wide. Spatial drift between the images of the stack was corrected using the method described in [38].

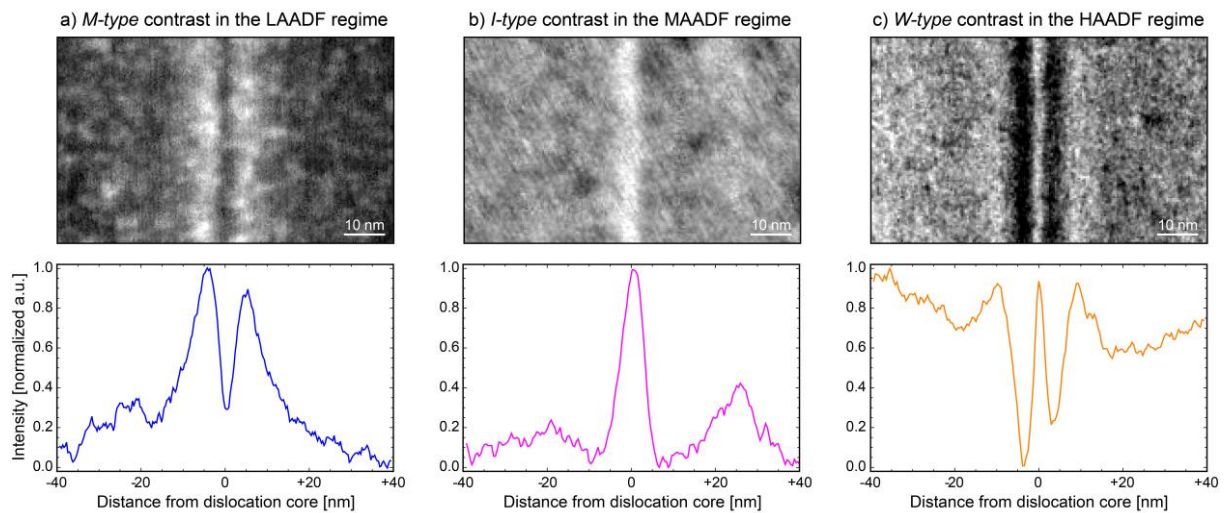

Figure 4. Experimental image contrast from a top-configuration dislocation at different ADF-STEM regimes: (a) LAADF regime, *M-type*; (b), MAADF regime, *I-type*; (c) HAADF regime, *W-type*. To the bottom of each image the corresponding intensity line profile is shown. The intensity profiles are integrated over 150 pixels.

The *M-type* contrast in the LAADF regime is directly interpretable using the classical diffraction Bragg theory [39]. Considering the significant contribution of the low-order elastic Bragg reflections to the contrast of the LAADF images, the appearance of the *M-type* contrast can be attributed to the changes in diffraction condition from the strained regions around the dislocation core. Due to local change of the deviation parameter ($s_g$) in the strained region, a positive contrast can be observed on either sides of the dislocation core. Due to contributions of multiple diffracted beams with a range of deviation parameters the region exhibiting positive contrast is wide (≈ 7 nm). The non-uniform intensity at either side of dislocation core is probably due to slight deviation from the exact zone axis condition.

De-channelling of the electron probe in the presence of strain has also been proposed as a source of enhanced signal in the LAADF regime [14]. In the de-channelling theory, changes in the electron wave vector due to the dislocation strain field result in spreading of the diffraction pattern. This is consistent with experimental observations in Figure 5, where we compare the CBED patterns of different regions across a dislocation line. The CBED pattern expands as the beam crosses the strained regions around dislocation core, and direction of this expansion is opposite at either sides of dislocation core. Therefore, it can be assumed that the origin of the intensity increase at either side of the dislocation core is due to the spanning of the expanded region of the CBED to the annular detector. This also explains why such enhanced signal disappears when the inner collection angle of the detector is larger than ≈ 40 mrad, e.g. for CL ≤ 330 mm, in the MAADF regime. However, interestingly no CBED spreading is observed at the core of dislocation and the corresponding CBED pattern resembles those of regions far away from dislocation strain field. This justifies why there is no gain in the LAADF signal from dislocation core, a region that accommodates the greatest atomic displacements. This latter phenomenon can be explained by taking into account that long-range strain and local static displacements affect the forward propagation of the wave function differently [17]. As proposed by Grillo et al., in case of large static displacements, i.e. regions around dislocation core, curvature of atomic planes should be taken into account as a source for the ADF signal [16]. Accordingly, the enhanced LAADF signal either side of dislocation core stems from the diffuse scattering (Huang scattering) that arises from the large lattice curvature in the strained regions around a dislocation core. In contrary, a trajectory passing through the core of an edge-character dislocation encounters a curvature field close to zero, thus no signal modification is expected from the region that corresponds to the core of dislocation; with the exception of a peculiarity at higher collection angles due to a minimal effect on the electron beam propagation that will be discussed later in Section 3.2.

The Huang scattering, produced by the bending of the atomic column, is mainly peaked in the forward direction, i.e. in the angular region between 20 and 40 mrad, due to the slowly changing displacement. Conversely, a very abrupt lateral displacement of the atoms occurs in the proximity of the dislocation core and its effects can be assimilated to the most disordered motions: thermal vibration. Following the idea proposed in the classic simple model [40], the scattering due to the presence of the extreme strain field at the dislocation core can be treated as the one related to the thermal diffuse scattering, therefore covering a larger angular range. There is therefore a regime from MAADF to moderate HAADF where this second type of Huang scattering (so-called extra Huang scattering) produces a positive contribution to the intensity in the *I-* and *W-type* contrast.

The Bloch-wave scattering theory can be used to further interpret the contrast features. Since the crystal is in a strong axial channelling condition (i.e. zone axis), electrons are well channelled before encountering a dislocation. Reduction of the *1s*-state due to crystal imperfection results in redistribution of intensity between Bloch states, producing Huang diffuse scattering in the LAADF regime. Excitation of the *1s*-state is linked directly to the generation of diffuse scattering, thus is the main effect responsible for signal at medium to high scattering angles. Difference in the scattering angle of the excited Bloch states has also be taken into account for the interpretation of the contrast. The scattering angle of the diffusely de-channelled electrons is not the same for all the Bloch states, and is larger for the *1s*-state compared to the other states. The reason is that the *1s*-state propagate very close to atomic sites thus, compared to other states that are mainly concentrated between the atoms, will experience diffuse scattering to higher angles. Therefore, it can be assumed that the scattering angle of the electrons being diffused from the dislocation core is greater than those generated due to strain around it, therefore can be discerned only by increasing detector inner collection angle to the MAADF regime. On the contrary, the strained regions at both sides of the dislocation core affect the excitation of higher Bloch states, therefore for larger inner collection angles, in the HAADF regime, these region appears with a negative contrast compared to the regions oriented in an axial channelling condition. As result, the contrast of a top-configuration dislocation in the HAADF regime is complementary to that of LAADF regime, i.e. the *W-type* contrast. Further increase of the outer collection angle increases the contribution from dynamic TDS to the image signal, resulting in overshadowing of the diffusely scattered signal from static atomic displacements, thus reduces the visibility of defects.

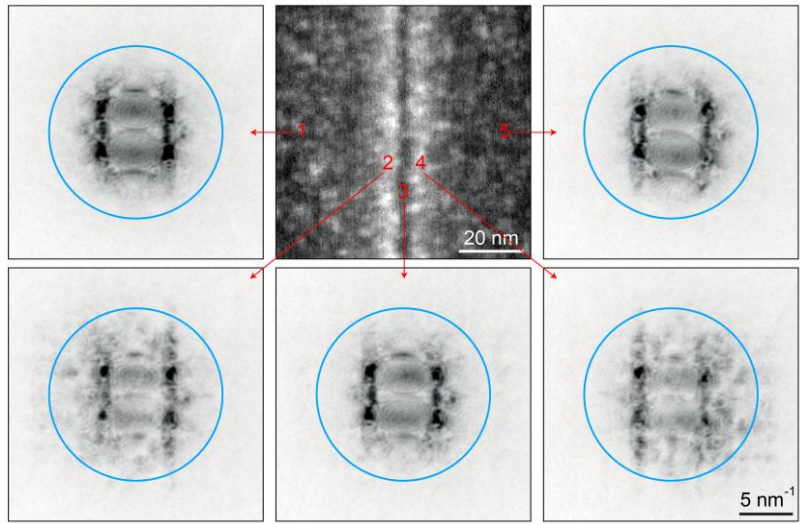

Figure 5 – Illustration of the energy-filtered CBED patterns from different regions around an edge-character dislocation. Expansion of the CBED pattern is obvious for the adjacent regions either side of dislocation core (regions 2 and 4). Relative to the intact regions (marked 1 and 5), the extent of the spreading of the CBED pattern is about 30 mrad, explaining the appearance of the *M-type* contrast in the LAADF regime only. The blue circle indicates the location of the annular detector of 20 mrad inner collection angle.

The depth-dependent contrast behaviour can also be described by diffuse scattering based on the Bloch-wave channelling model and is regarded to be due to the attenuation of the *1s* Bloch state in the specimen thickness. Channelling related effects that give rise to dislocation contrast at medium to high angles, are less important in the bottom of a thick specimen, hence dislocations located at the bottom segment should appear with a faint contrast. On the other hand, the enhanced signal in the *M-type* contrast of the LAADF regime, is caused by the de-channelling of any Bloch state, and not only the *1s* state. Since high-energy states remain populated even at the bottom of the foil, the LAADF signal is less sensitive to the position of the dislocation in the foil. Conversely, the *I-type* and *W-type* contrasts, dominated by scattering of the *1s* Bloch state, fades for the bottom configuration dislocation in the MAADF and HAADF regimes.

### 3.2. Computational results

To provide a computational validation of the proposed theory, numerical simulations using Grillo's equation were performed. Taking into account all the factors that influence the electron beam propagation inside the sample, the ADF dislocation contrast has been computed for three different angular regimes (i.e. LAADF, MAADF, and HAADF) to cover the different experimental settings.

As expected, depending on the strain amplitude, the atomic columns around dislocation core are distorted differently. When the curvature is too high the material is to all effects amorphous and no change in the *1s*-state is expected, as demonstrated in Figure 7, where a series of multislice simulations of the propagation of the electron beam along a curved atomic column is shown. The sample represents a single Ga column in a 15 nm thick GaN foil with the $[1\bar{1}00]$ direction oriented along the electron beam direction. The electron beam propagation has been simulated for different curvature radii. For small curvature radii (Figures 7b-e), the electron beam is able to follow the curved column. As a result of the curvature, however, the beam is rapidly de-channelled as testified by the rapidly decreasing intensity and the disappearing of the *pendellösung* oscillations. Instead, when the curvature exceeds a critical value (Figure 7f), the electron channelling is completely disrupted and the beam propagates as it does in an amorphous material. In order to take into account the effect if large atomic column curvatures, the curvature equation has been updated to the following expression:

$$\Gamma' = \begin{cases} \frac{m2}{\pi} sin\left(\frac{\pi\Gamma}{2m}\right) & if\ |\Gamma| < 2m \\ 0 & otherwise \end{cases}$$

This approach was also tacitly used in the previous Grillo's equation version [16], but here we made it more explicit. According to this equation, for very large strains, the atomic column curvature can be considered close to zero, meaning that the beam does not follow the curvature. Therefore, a HAADF signal similar to that of the unstrained matrix is expected at the dislocation core. However, as can be seen in the experimental images, in the dislocation centre where $\Gamma'$ is supposed to be 0, the intensity is larger than in the matrix. The MAADF intensity at the core location also shows a similar behaviour. To explain this phenomenon we go in more details of the Huang scattering main features. As explained earlier, we are facing a second type of diffuse scattering (i.e. extra Huang scattering) that is qualitatively different from the one described in the previous Grillo's strain-channelling equation. Such diffuse scattering covers moderate angles (MAADF and HAADF regimes), and we expect that its contribution should however disappear in the HAADF regime with very large collection angle, as evidenced by experimental images.

In ADF simulations using Grillo's strain-channelling equation, the resulting contrast is highly dependent on diffusion related parameters as well as on the contribution of Huang scattering. By properly adjusting these parameters, the contrast features in the three regimes were simulated, and show a good agreement with the experimental images, as demonstrated in Figure 8. The *I*-type contrast however appears slightly wider in the experimental images. This is because the scattered electrons can undergo thermal re-scattering as they propagate into the crystal.

The depth-dependent ADF contrast behaviour has been further investigated. Computed LAADF, MAADF, and HAADF profiles for a dislocation placed at different depth are demonstrated in Figure 9. The LAADF profile does not show any remarkable change with dislocation position, and appears nearly the same for the whole range of dislocation depths. In the MAADF, the profile changes from *I-type* to *M-type* as the dislocation depth is increased; i.e. change from an enhanced signal at the core reduced to signal peak at either side of dislocation core. This is consistent with experimental observations, further validating the proposed theory based on the attenuation of the *1s* channelling in the foil depth. Similarly, in the HAADF regime the contrast from dislocation gradually decreases by increasing the depth of dislocation in the foil and a bottom-configuration dislocation is barely visible in this regime. Overall, the close agreement between the experimental and computational images corroborates the proposed model for interpreting the dependence of ADF dislocation contrast to its depth and detector collection range.

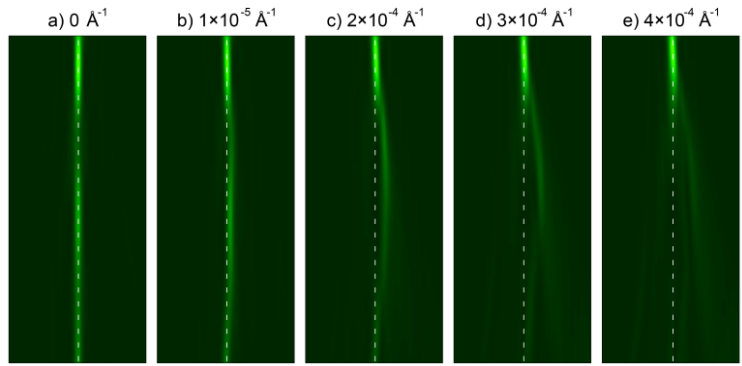

Figure 6. Multislice simulation showing the effect of atomic column curvature on the electron beam propagation along a 15 nm thick GaN atomic column. The nominal atomic column curvatures are a) 0 Å$^{-1}$, b) 1×10$^{-5}$ Å$^{-1}$, c) 2×10$^{-4}$ Å$^{-1}$, d) 3×10$^{-4}$ Å$^{-1}$, e) 4×10$^{-4}$ Å$^{-1}$. Dashed lines represent the intact atomic column.

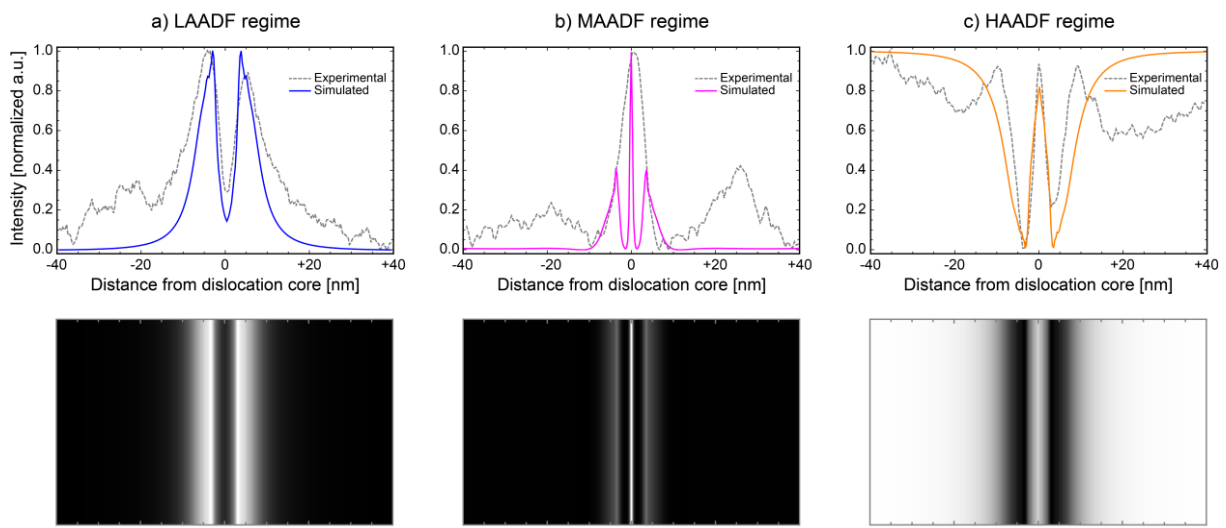

Figure 7. Comparison between the experimental and simulated ADF-STEM images of a top-configuration dislocation. The top row shows the intensity profiles across the experimental and simulated images at different ADF regimes: a) LAADF regime, *M-type*; b) MAADF regime, *I-type*; and c) HAADF regime, *W-type*. Simulated images are presented in the bottom row.

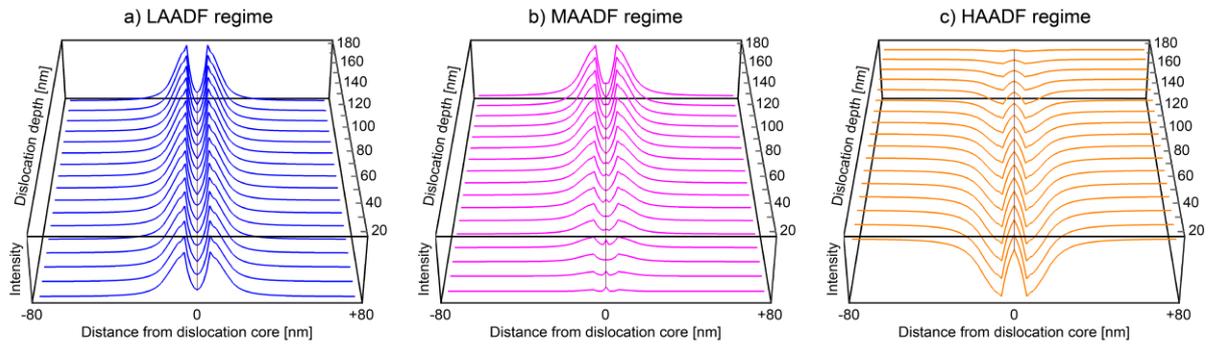

Figure 8. Evolution of the ADF-STEM dislocation contrast as a function of dislocation depth in the specimen. Intensity profiles across the simulated images from edge-character dislocations of different specific depths in the foil are plotted for all three ADF regimes: a) LAADF regime, b) MAADF regime, and c) HAADF regime.

## 4. Conclusions

In summary, ADF-STEM image contrast from dislocations in a GaN membrane was systematically studied. Experimental images showed that contrast features are sensitive to the collection range of the ADF detector as well as to the specific depth of the dislocation in the specimen. For a dislocation located close to the entrant foil surface, strain-induced Huang scattering, along with other mechanisms such as de-channelling and redistribution of the Bloch state population by dislocation, cause signals scattered to different angles. Therefore, depending on the ADF detector collection range different contrasts are obtained: the *M*-, *I*-, and *W-type* contrasts in the LAADF, MAADF, and HAADF regimes, respectively. All the *s*-state related mechanisms, which are believed to be the origin of a positive contrast from the dislocation core in the MAADF and HAADF images, peter out with the distance into the crystal depth, thus the *I*- and *W-type* contrasts are not observed for dislocations situated beyond a certain depth. A rationale based on the electron channelling and Bloch-wave scattering was proposed to account for the mechanisms underlying the experimental contrast features. Grillo's strain-channelling equation was implemented for the simulations of the ADF-STEM dislocation contrast to provide a computational validation of the proposed mechanisms.

**Acknowledgments**

E.O would like to acknowledge: Prof. Pierre Stadelmann, Dr. Duncan T.L Alexander, Dr. Quentin Jeangros of CIME-EPFL, Prof. Peter Nellist of Oxford University, and Prof. Michael Mills of The Ohio State University for fruitful discussions; Prof. Nicolas Grandjean of LASPE-EPFL for supplying the specimen; Daniele Laub of CIME-EPFL for providing help with TEM sample preparation. Swiss National Science Foundation financially supported this work (project no. 200020-143217). E.R. and V.G. acknowledge the support from Q-SORT, a project funded by the European Union's Horizon 2020 Research and Innovation Program under grant agreement No. 766970.

# Supporting information

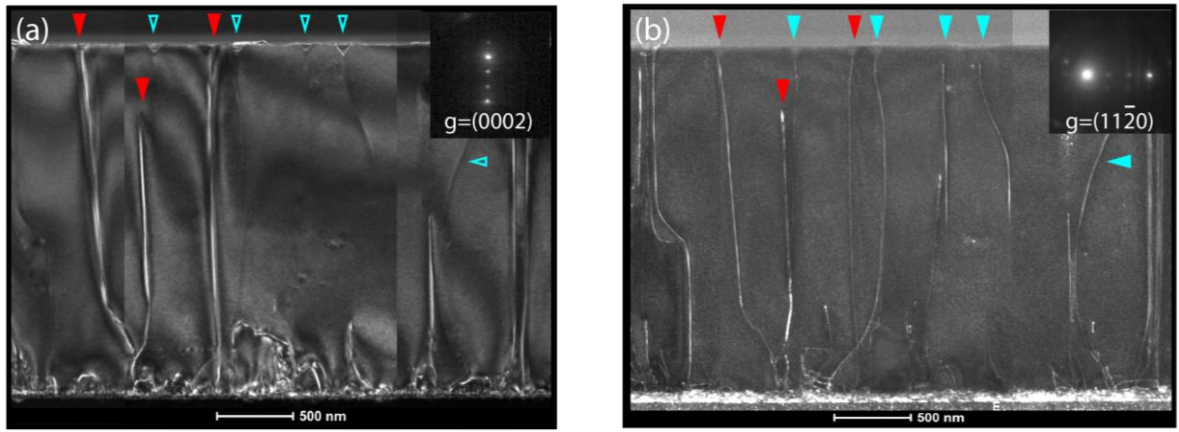

Supplementary Figure S1. Dislocation Burgers vector analysis in the GaN layer. Weak-beam dark-field ($\mathbf{g} - 3\mathbf{g}$) CTEM images with a) $\mathbf{g} = (0\,0\,0\,2)$ and b) $\mathbf{g} = (1\,1\,\bar{2}\,0)$ operative diffraction vectors (close to $[1\,\bar{1}\,0\,0]$ zone axis). Corresponding diffraction pattern for each image is shown in the inset. The edge and mixed type dislocations are indicated by cyan and red triangles, respectively: filled, visible on the image; empty, invisible on the image. Edge dislocations, having the Burgers vector $\mathbf{b} = \frac{a}{3}\langle 1\,1\,\bar{2}\,0\rangle$, are invisible under $\mathbf{g} = (0\,0\,0\,2)$ diffraction condition. Contrast of the screw dislocations with $\mathbf{b} = a\langle 0\,0\,0\,1\rangle$ disappear on the images obtained with $\mathbf{g} = (1\,1\,\bar{2}\,0)$ in (b). Mixed dislocations with $\mathbf{b} = \frac{a}{3}\langle 1\,1\,\bar{2}\,\bar{3}\rangle$ are visible under both imaging conditions.

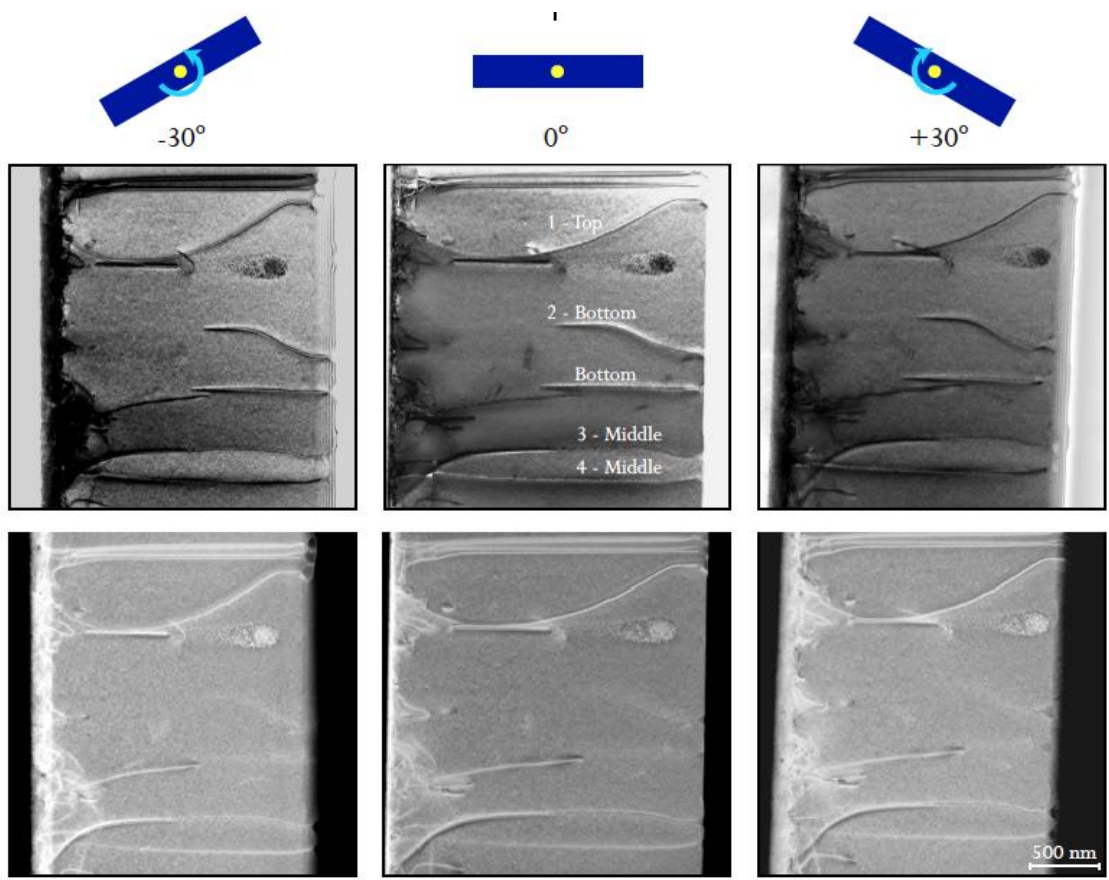

Supplementary Figure S2 – Bright-field STEM images (top row) of dislocations in a GaN lamella with its [1 $\bar{1}$ 0 0] direction parallel to the electron beam (centre), and tilted to -30° and +30° degree (left and right) along the **g** = (0 0 0 2) direction. Intersection of a dislocation line and thickness fringes can provide a rough idea about the specific depth of the dislocation in the foil. In this configuration, the first thickness fringe corresponds to the foil's top for the images of negative tilt direction. In the bottom row, corresponding HAADF-STEM images are demonstrated, highlighting the depth-dependent contrast of dislocations in the HAADF regime; the visibility of the dislocation decreases with its specific depth in the foil.

# Appendix

This upgraded model directly ties to the classic simple model for the Huang scattering where an additional static Debye Waller factor M' is added to the normal thermal factor M [40]. This effect can be added to the cross section $\sigma$ that is:

$$\Delta\sigma \propto \int_{\theta_{min}}^{\theta_{max}} f^2(\theta)\left[1 - e^{-2(M+M')(\theta^2/\lambda^2)}\right]d^2\theta - \int_{\theta_{min}}^{\theta_{max}} f^2(\theta)\left[1 - e^{-2M(\theta^2/\lambda^2)}\right]d^2\theta$$

, where $f$ is the elastic scattering factor, $\theta$ is the scattering angle, $\lambda$ is the electron beam wavelength, $M$ is the Debye Waller and $M'$ is the additional Debye Waller factor that takes into account the effect due to disorder. Moreover, such additional Debye Waller factor produces a more rapid absorption of $1s$ states but is no considered here. Another less fundamental approach is to define a corrected atomic column curvature $\Gamma''=\Gamma-\Gamma'$. Taking into account the above introduced factors, the previously introduced Grillo's set of strain-channelling equations can be written as:

$$\begin{cases} \dfrac{d^2\Phi_{1s}(x,y,z)}{dz^2} = \left[-2((\mu_{1s}+\mu_{SD}))\dfrac{d\Phi_{1s}(x,y,z)}{dz} - 2\left(\dfrac{\Gamma^2}{\sigma_\theta^2}\right)\Phi_{1s}(x,y,z)\right] \\ \dfrac{dI_{Huang}(x,y,z)}{dz} = 2H\left(\dfrac{\Gamma^2}{\sigma_\theta^2}\right)\Phi_{1s}(x,y,z) \\ I = \int(\Phi_{1s}+C)\left[\sum_{i=atoms}\left(\sigma_i + H'\left(\dfrac{\Gamma''^2}{\sigma'_\theta^2}\right)\right)\delta(z-z_i)\right]dz + I_{Huang} \end{cases}$$

, where $\Phi_{1s}$ is the excitation of the $1s$ Bloch states, (x,y,z) are the space direction, $\mu_{1s}$ is the absorption of the $1s$ Bloch states coefficient, $\mu_{SD}$ is the absorption coefficient related to the static displacement, $\sigma_i$ is the scattering cross section for the $i_{th}$ atom in the relevant atomic column, $\sigma_\theta$ is the size of the $1s$ state in the reciprocal space, $I_{Huang}$ is the Huang scattering contribution, $C$ is an adjustable parameter to account for the non-$1s$ states contribution, $I$ is the ADF intensity and $H'$ is the parameter that takes into account of the extra Huang scattering.